\documentclass[aps,prd,preprint,superscriptaddress,showpacs,showkeys]{revtex4}
\usepackage{amssymb,amsmath}
\usepackage{graphicx} % Include figure files
\usepackage{dcolumn}  % Align table columns on decimal point
\usepackage{epsfig}   % For PostScript figures
\usepackage{pstricks}
\usepackage{ifpdf}
\newcommand{\tr}{ {\mathrm{tr}\, }}
\newcommand{\Tr}{ {\mathrm{Tr}\, }}

\begin{document}

% Use the \preprint command to place your local institutional report
% number in the upper righthand corner of the title page in preprint mode.
% Multiple \preprint commands are allowed.
% Use the 'preprintnumbers' class option to override journal defaults
% to display numbers if necessary
%\preprint{}
%\preprint{INLN May 2012}

%Title of paper
\title{Casimir operator dependences of non-perturbative fermionic QCD amplitudes}

% repeat the \author .. \affiliation  etc. as needed
% \email, \thanks, \homepage, \altaffiliation all apply to the current
% author. Explanatory text should go in the []'s, actual e-mail
% address or url should go in the {}'s for \email and \homepage.
% Please use the appropriate macro foreach each type of information

% \affiliation command applies to all authors since the last
% \affiliation command. The \affiliation command should follow the
% other information
% \affiliation can be followed by \email, \homepage, \thanks as well.
%
%\author{T. Grandou}
%\affiliation{Universit\'{e} de Nice-Sophia Antipolis,\\ Institut Non Lin\'{e}aire de Nice, UMR 6618 CNRS 7335; 1361 routes des Lucioles, 06560 Valbonne, France}
%\email[]{Thierry.Grandou@inln.cnrs.fr}
%
%\author{H. M. Fried$^{\dagger}$, T. Grandou$^\star$, Y.-M. Sheu$^{\star}$}
%\affiliation{${}^\dagger$ {Physics Department, Brown University, Providence, RI 02912, USA} \\ ${}^\star$ {Universit\'{e} de Nice-Sophia Antipolis, Institut Non Lin$\acute{e}$aire de Nice, UMR 6618 CNRS 7335; 1361 routes des Lucioles, 06560 Valbonne, France}}
%\email[]{Your e-mail address}
%\homepage[]{Your web page}
%\thanks{}
%\altaffiliation{}
%\affiliation{Physics Department, Brown University, Providence, RI 02912, USA}

\author{H.M. Fried}
\affiliation{Physics Department, Brown University, Providence, RI 02912, USA}
\email[]{Herb Fried fried@het.brown.edu}
\author{T. Grandou}
\affiliation{Universit\'{e} de Nice-Sophia Antipolis,\\ Institut Non Lin\'{e}aire de Nice, UMR CNRS 7335; 1361 routes des Lucioles, 06560 Valbonne, France}
\email[]{Thierry.Grandou@inln.cnrs.fr}
\author{R. Hofmann}
\affiliation{Institut f\"ur Theoretische Physik\\ 
Universit\"at Heidelberg\\ 
Philosophenweg 16\\ 
69120 HEIDELBERG}
\email[]{r.hofmann@thphys.uni-heidelberg.de}

%Collaboration name if desired (requires use of superscriptaddress
%option in \documentclass). \noaffiliation is required (may also be
%used with the \author command).
%\collaboration can be followed by \email, \homepage, \thanks as well.
%\collaboration{}
%\noaffiliation

\date{\today}

\begin{abstract} In eikonal and quenched approximation, it is argued that the strong coupling fermionic QCD Green's functions and related amplitudes depart from a sole dependence on the $SU_c(3)$ quadratic Casimir operator, $C_{2f}$, evaluated over the fundamental gauge group representation. \par
Noticed in non-relativistic Quark Models and in a non-perturbative generalization of the Schwinger mechanism, an additional dependence on the cubic Casimir operator shows up, in contradistinction with perturbation theory and other non-perturbative approaches. However, it accounts for the full algebraic content of the rank-2 Lie algebra of $SU_c(3)$. Though numerically sub-leading effects, cubic Casimir dependences, here and elsewhere, appear to be a signature of the non-perturbative fermonic sector of QCD.\end{abstract}

% insert suggested PACS numbers in braces on next line
% insert suggested keywords - APS authors don't need to do this

%\maketitle must follow title, authors, abstract, \pacs, and \keywords
\maketitle

% body of paper here - Use proper section commands
% References should be done using the \cite, \ref, and \label commands
\section{\label{SEC:1}Introduction}
% Put \label in argument of \section for cross-referencing
%\section{\label{}}
%\subsection{}
%\subsubsection{}

In some recent articles~\cite{QCD1,QCD-II,QCD5, QCD6, QCD5'}, a property, which bears on the non-perturbative fermionic Green's functions of QCD, has been put forth under the name of {\textit{effective locality}}. This property
can be phrased as follows. For any Quark/Quark (or Anti-Quark) scattering amplitude, the
full gauge-invariant sum of cubic and quartic gluonic interactions,
fermionic loops included, results in a local contact-type interaction, and this local interaction is
mediated by a tensorial field which is antisymmetric both in Lorentz and
color indices. This is a non-expected result because, ordinarily, integrations of elementary degrees of freedom result in highly non-local structures;  the
`effective locality' denomination, which sounds like an {\textit{oxymoron}}, accounts for this
rather unusual circumstance. It is worth emphasizing that effective locality is an exact property of full QCD, and that its derivation entails no approximation \cite{QCD-II}.

Then the consequences of effective locality, even when examined `at tree level', should exhibit admissible as well as new aspects of the confined phase of QCD; and so far, it seems to be so \cite{QCD-II, QCD5, QCD6, QCD5'}.

In Ref.\cite{QCD6}, a general form of the non-perturbative QCD fermionic amplitudes is displayed as a finite sum of finite products of {\textit{Meijer special functions}}, in agreement with general expectations~\cite{Ferrante2011}. Remarkably enough, within one and the same expression, these amplitudes are able to display an explicit link between a {\textit{partonic content}} and a hadronic non-perturbative component in accord, this time, with the $AdS_5/QCD$ light-cone approach of Ref.\cite{deTeramond2012}.

However, the analysis presented in Ref.\cite{QCD6} is carried out at eikonal and quenched approximations. Soon it will become important to relax these approximations, not only for the sake of preserving unitarity, but also in order to explore larger distances: Effective locality, in effect, clearly differentiates QCD from a pure Yang Mills situation. In particular, as noticed also by lattice approaches, inclusion of quark loops reveals to be essential to the description of larger distance non-perturbative physics~\cite{QCD5}.

Fortunately, some things can be learnt already at the level of a quenched (and eikonal) analysis. In a recent letter \cite{tg}, it was argued that non-perturbative fermionic Green's functions, and amplitudes thereof, do not only depend on the quadratic Casimir invariant, $C_2$, but also on the trilinear Casimir invariant, $C_3$.
\par
This extra dependence obviously complies with the rank-2 character of an $SU_c(3)$-color algebra, and at this level of approximations at least, it is a peculiar output of effective locality. With the exception of Ref.\cite{{Nieuwenhuizen}, {Cooper}} some years ago, it is remarkable that $C_3$ has come unnoticed for so long, whereas its importance was put forth in nonrelativistic quark models \cite{Dmitrasinovic}. Now, the relative smallness of $C_{3f}$-dependences, as displayed in Section IV, is certainly at the origin of this fact.

\par
Be it as it may, it will certainly matter to disentangle such a prediction from the approximated context where it was first discovered \cite{prep}. At present, though, the current paper aims at making more complete and accurate the non-trivial results first announced in Ref.\cite{tg} and display some numerical estimates.

 \par\medskip
The paper starts from an expression for the fermonic $4$-point function which is the matter of a whole article \cite{QCD6}. For the sake of traceability, Appendix A offers a summary of the main steps at the origin of that expression. Besides, the paper is organized as follows. \par
For an easier and explicit presentation, Section II introduces the matter in the case of a $4$-point fermionic Green's function, while providing with Appendix D, the necessary extension to the general case of $2n$-point Green's functions. In order to alleviate a presentation which is already quite technical, some proofs are deferred to Appendices B, D, E and F. \par Section III deals with the necessary average to be taken over the orthogonal group of matrices $O_N(\mathbb{R})$, the procedure which reveals the full color algebraic structure of $2n$-point fermonic  Green's functions and related amplitudes. A proof necessary to Section III is produced in Appendix C, and a conclusion is proposed in Section V.

\section{\label{SEC:2} Non-Perturbative fermionic Green's Functions}

In Perturbation Theory, all of the scattering process calculations come out proportional to either $C_A=N_c$ and/or $C_F={(N_c^2-1)}/{2N_c}$, that is to the quadratic Casimir operator eigenvalue $C_2({\cal{R}})$ over the adjoint and fundamental representations respectively (exceptions may be found in the lattice-gauge theory approach of Ref.~\cite{Bali}, where higher dimensional representation spaces were considered, but again, restricted to $C_2$-dependences). The quadratic Casimir operator's well known definition is $C_2({\cal{R}})=\sum_{a=1}^{8} T_a^2({\cal{R}})$ where the $T_a({\cal{R}})$ denote the $SU_c(3)$ Lie algebra generators in a given representation ${\cal{R}}$.
 \par
Not only perturbative calculations, but also non-perturbative QCD models, such as the MIT bag model~\cite{MIT}, the Stochastic Vacuum Model (SVM)~\cite{Dosch}, and lattice approaches~\cite{Bali,BMuller}, comply with these overall $C_2({\cal{R}})$ dependences, though, sometimes, it is worth noticing, in quite different ways. It could be that, indeed, perturbative inputs be imported as surreptitious elements, from perturbation theory to these non-perturbative attempts \cite{BMuller}.

The property of  effective locality surfaces at the level of non-perturbative $2n$-point fermionic Green's functions, as an exact, non-approximate property of QCD~\cite{QCD-II}. In a strong coupling regime of $g>>1$, evaluating a 2 to 2 Quark(Anti-Quark)/Quark(Anti-Quark)  scattering amplitude with the help of (analytically continued \cite{QCD6}) {\textit{Random Matrix Theory}} \cite{Mehta1967}, one finds a result proportional to~\cite{QCD6},
\begin{eqnarray}\label{Eq:1}
&& (-16\pi{m^2\over E^2})^N\sum_{\mathrm{monomials}}(\pm 1)\,\Tr\prod^{\sum q_i=N(N-1)/2}_{1\leq i\leq N}\ [1-i(-1)^{q_i}] \nonumber\\  & & \!\!  \times\  C\int{\rm{d}}p_1\  ..\ {\rm{d}}p_{N(N-1)/2}\ f(p_1,\dots, p_{N(N-1)/2})\,\! \int_0^{+\infty} {\rm{d}}\alpha_1^i\ {\sin[\alpha_1^i({\cal{OT}})_i]\over \alpha_1^i}\int_0^{+\infty} {\rm{d}}\alpha_2^i\ {\sin[\alpha_2^i({\cal{OT}})_i]\over \alpha_2^i} \nonumber\\  & & \quad \times \, G^{23}_{34}\left( \left.{iN_c}\left({ \alpha_1^i\alpha_2^i \over g\varphi(b) }\right)^2{{\hat{s}}({\hat{s}}-4m^2)\over 2m^4} \right|
%\matrix{{(3-2q_i)/ 4},&{1/ 2},&1\cr {1/ 2},& {1/ 2},&1,&1\cr}
\begin{array}{cccc}
  \frac{3 - 2 q_i}{4}, & \frac{1}{2}, & 1, &  \\
  1, & 1, &  \frac{1}{2}, &  \frac{1}{2}
\end{array}
\right)\, ,
\end{eqnarray}where the eikonal and quenching approximations have been used. QCD is here simplified to the case of a single quark species of mass $m$, and $E(=E_1=E_2)$ is each of the two colliding quarks energy in the center of mass system, $p_1=(E,0,0,p)$, $p_2=(E,0,0,-p)$, ${\hat{s}}=(p_1+p_2)^2$. In Appendix D (and thanks also to Appendix F), it is shown how this $4$-point expression  generalizes to the case of $2n$-point fermionic Green's functions. \par
 Note that expression (\ref{Eq:1}) is the matter of a whole paper \cite{QCD6}. For the sake of providing it with enough background, though, the main steps of its derivation are summarized in Appendix A.
\par\medskip\noindent
The variables entering (\ref{Eq:1}) are the following.
\par\medskip
- In (\ref{Eq:1}), ${\cal{O}}={\cal{O}}(.., p_j,..)$ stands for an orthogonal $N\times N$ matrix specified by the $N(N-1)/2$ parmeters $p_js$, with $N=D\times(N_c^2-1)$, $D$, the number of space-time dimensions; that is $N=32$, \cite{{Halpern1977a},{Halpern1977b}}. The distribution $f(.., p_j,..)$ defines the {\textit{Haar measure}} of integration over the orthogonal group ${O_N(\mathbb{R})}$, and the constant $C$, its normalization ({\textit{i.e.}}, $C$ is the inverse of the ${O_N(\mathbb{R})}$-volume).  
\par\smallskip
- In (\ref{Eq:1}), the $N\times N$ orthogonal matrices ${\cal{O}}$ act upon an $N$-vector of matrices ${\cal{T}}=(1,1,1,1)\otimes T=(T,T,T,T)$. That is, ${\cal{T}}$ is made out of $D=4$ copies of the full set $T$ of $SU_c(3)$ generators, taken in the fundamental representation: $T=\{t_1,t_2,\dots,t_7,t_8\}$, with $t_a={\lambda_a/2}$, the standard Gell-Mann matrices~\cite{Yndurain}. However technical, this step, related to (\ref{M}), allows one to perform a series of exact integrations in a systematic way.
\par\smallskip
- The third line of (\ref{Eq:1}) displays a {\textit{Meijer special function}}, $G^{23}_{34}$. How these Meijer functions come into play is displayed by (\ref{Meijer}), Appendix A. The Meijer function depends on an array of parameters, one of them involving an integer $q_i$, with $1\leq i\leq N$ and $0\leq q_i\leq N(N-1)/2$. In Appendix A, Equation (\ref{VdM}), it is shown that this power of $q_i$ (see also (\ref{Meijer}), at $p=q_i-1/2$) comes from the expansion of a {\textit{Vandermonde determinant}} into a sum of monomials. It is this sum which is explicitly referred to in the first line of (\ref{Eq:1}).

 In one and the same argument, it is worth pointing out that the $G^{23}_{34}$ Meijer function's argument mixes up partonic variables ($m, {\hat{s}}$) with the non-perturbative function,
 \begin{equation}\label{phi} \varphi(b)=({\mu/{\sqrt{\hat{s}}}})\,e^{-(\mu b)^{2-\xi}}\,,\end{equation} where $b=|{\vec{b}}|$ is the impact parameter of the 2 scattering quarks in their center of mass system, and $\mu$, the mass scale necessarily introduced by the property of effective locality \cite{QCD6}, as stated in Appendix A, after Equation (\ref{currents}). This is interesting because, as quoted in the Introduction, that point turns out to be a prediction of the $AdS_5/QCD$ approach to the non-perturbative regime of QCD \cite{deTeramond2012}. In (\ref{phi}), $\xi$ stands for a small {\textit{deformation parameter}}, on the order of one tenth \cite{QCD-II}. While this parameter is crucial to a description of confinement in terms of a linear potential between quarks \cite {{QCD1},{QCD-II}, {QCD5}, {QCD6}}, it has no relevance at all to the current considerations. Likewise, the non-perturbative function of (\ref{phi}) to be discussed in details elsewhere, is partly phenomenological; this point, however, has only a very marginal incidence on the results of the present paper.
\par\medskip
Two remarks are in order.
\par
(i) Deriving (\ref{Eq:1}), the absolute values of (\ref{measure}), relevant to the QCD case of $\kappa=1$, have been dropped. That this simplification can be made without compromising the overall color algebraic structure of Green's functions, is shown in Appendix B.  

\par\medskip
(ii) The color algebraic part of (\ref{Eq:1}) appears in its 2nd line in terms of the 2 {\textit{sine functions}} with their arguments being the $N_c\times N_c$ matrices $[\alpha_J^i({\cal{OT}})_i]\,,\ J=\{1,2\}$. For each monomial of the Vandermonde determinant (\ref{VdM}), the structure is that of a finite product of terms attached, each, to a given value of the index $i$, ranging between $1$ and $N$, and no summation understood on $i$.  At this order of approximations at least, this amounts to the statement that fermonic Green's functions and related amplitudes split into a finite sum of finite products of Meijer special functions. \par
 As will be further commented, this simple structure is far from being a trivial point, as it could be connected to the deep meaning of effective locality. In Appendix F it is shown that {\textit{Baker-Campbell-Hausdorf formulae}} would definitely object to it. However, this structure emerges as a consequence of the $O_N(\mathbb{R})$ average to be taken in (\ref{Eq:1}).

\par\bigskip
Now, as shown in Appendix E, it is possible to rely on a series of identities and exact textbook integration {\textit{formulae}}, so as to take (\ref{Eq:1}) to the form, 
\begin{eqnarray}\label{new1}
&& \nonumber\pm  (-{16\pi^2 m^2\over E^2})^N\sum_{\mathrm{monomials}}\biggl\langle\,\prod_{i=1}^N\, [1-i(-1)^{q_i}] \\ &&\nonumber\times\, \biggl[{{\sqrt{2iN_c}}\,{\sqrt{{\widehat{s}}({\widehat{s}}-4m^2)}}\over {m^2}}\biggr]\, \frac{[({\cal{OT}})_i]^{-2}}{g\varphi(b)}\\ &&\times\, G^{30}_{03}\!\left( \biggl[{  g\varphi(b)\over {\sqrt{32iN_c}}  }{m^2\over {\sqrt{{\widehat{s}}({\widehat{s}}-4m^2)}}}\biggr]^2\biggl[({\cal{OT}})_i\biggr]^{4}\, \biggr|\frac{1}{2}, \frac{3+2q_i}{4},1\!\right)\,\biggr\rangle_{\!O_N(\mathbb{R})}\ ,
\end{eqnarray}
where the $G^{30}_{03}$ Meijer function is explicitly defined in (\ref{Meijer}). In (\ref{new1}), the large brackets are here to denote the orthogonal group $O_N(\mathbb{R})$-average specified in the 2nd line of (\ref{Eq:1}). In random matrix calculations, it often happens that the eigenvalue spectrum only matters, and that the $O_N(\mathbb{R})$ averages just factor out and disappear in the normalization. This is not the case here: As pointed out in Refs.\cite{QCD6, tg} in effect, the additional but unavoidable complexity coming from the $O_N(\mathbb{R})$-average is essential not only to prevent a trivial result from occuring, but also to display the full algebraic content of the rank-2 $SU_c(3)$ color algebra. 
\par
The matrix-valued argument of the Meijer's function of (\ref{new1}) is,
\begin{equation}\label{z_i}
z_i\equiv\lambda\, [({\cal{OT}})_i]^{4}\,,\ \ \ \ \ \ \lambda\equiv \left( \,{  g\varphi(b)\over {\sqrt{32iN_c}}  }{m^2\over{\sqrt{{\widehat{s}}({\widehat{s}}-4m^2)}}}\,\right)^2\,.
\end{equation} As argued in Appendix E, even at a sufficiently  large value of the coupling constant $g$, the matrix elements of $z_i$ are much smaller than unity, each. It is therefore possible to rely on the standard Meijer function expansion, (\ref{D5}), (\ref{D6}), and write for a given monomial appearing in the sum (\ref{new1}),
\begin{equation}\label{zexpand}
\pm  (-{4\pi^2 m^2\over E^2})^N\biggl\langle\prod_{i=1}^N\, [1-i(-1)^{q_i}]\,\sum_{h=1}^3A^h_i\,z_i^{b_h-\frac{1}{2}}\left(\,1+O_{ih}\,z_i+{\cal{O}}(z_i^2)\,\right)\biggr\rangle_{\!O_N(\mathbb{R})}\,,
\end{equation}
where the $b_h$ stand for the parameters of the $G$-Meijer function of (\ref{new1}), that is $b_h=\{b_1,b_2, b_3\}=\{1/2,(3+2q_i)/4, 1\}$. Out of Appendix E, numbers $A^h_i$ are readily identified to be given by,
\begin{equation}\label{Anumbers}
A^1_i=\Gamma(\frac{1}{2})\Gamma(\frac{2q_i+1}{4}),\  \ \ \ A^2_i=\Gamma(\frac{1-2q_i}{4})\Gamma(\frac{-2q_i-1}{4}),\  \ \ \ A^3_i=\Gamma(-\frac{1}{2})\Gamma(\frac{2q_i-1}{4})\,,
\end{equation}and likewise, 
\begin{equation}\label{Onumbers} O_{i1}=\frac{8}{2q_i-3}, \  \ \ \ O_{i2}=\frac{16}{(2q_i+3)(2q_i+5)},\    \ \ \ O_{i3}=-\frac{8}{2q_i+3}\,.
\end{equation} Note that, with no incidence at all for the sequel, (\ref{zexpand}) brings a correction to Eq.(8) in Ref.\cite{tg}.

\section{\label{SEC:3} $O_N(\mathbb{R})$-integration and Casimir operator dependences}

Thanks to effective locality, the original (and infinite dimensional) $\chi^a_{\mu\nu}$-functional summations of (\ref{Halpern}),
 \begin{equation}\label{dchi}
\int{\mathrm{d}[\chi]} = \prod_{w\in\mathcal{M}} \prod_{a=1}^{8} \prod_{ 0= \mu<\nu}^3 \int {\mathrm{d}[\chi_{\mu \nu}^{a}(w)]}
\end{equation} 
can be translated into an analytically continued Random Matrix integration \cite{QCD6} (in (\ref{dchi}), ${\cal{M}}$ is the four dimensional spacetime manifold). The latter splits into an integration on the spectrum of matrices $M$ defined in (\ref{M}), and an integration over the orthogonal group $O_N(\mathbb{R})$. As displayed in (\ref{Meijer}), the former yields the Meijer functions, whereas the full color algebraic dependences of fermionic Green's functions come about as the $O_N(\mathbb{R})$-integration is carried through. 
\par\medskip
The matrix-valued argument $z_i$ is on the order of $[({\cal{OT}})_i]^{4}$, and thus, the orders $z_i^0$, ${\sqrt{z_i}}$, $z_i$ and $z_i{\sqrt{z_i}} $ contributions of (\ref{zexpand}), are of even orders $({\cal{OT}})_i^0$, $({\cal{OT}})_i^2$, $({\cal{OT}})_i^4$ and $({\cal{OT}})_i^6$ respectively. They are leading contributions in view of the smallness of $\lambda$ (Appendix E), whereas odd powers of $({\cal{OT}})_i$s, such as $2q_i+1$, vanish trivially under $O_N(\mathbb{R})$-averaging. Random orthogonal matrices, ${\cal{O}}$, can be generated in different ways, distributed according to the {\textit{Haar measure}} over the orthogonal group $O_N(\mathbb{R})$~\cite{Anderson et al.}. An orthogonal matrix is conveniently decomposed into a product of $N(N-1)/2$ {\textit{rotators}}, plus reflections, 
\begin{equation}\label{O}
{\cal{O}}=(R_{12}R_{13}\dots R_{1N})\,(R_{23}R_{24}\dots R_{2N})\dots \dots (R_{N-1,N})\,D_{{\mathbf{\varepsilon}}}\,,
\end{equation}
where the matrix of reflections is diagonal by definition and reads, $D_\varepsilon=diag(\varepsilon_1, \varepsilon_2,\dots,\varepsilon_N)$, with $\varepsilon_i=\pm 1,\ \forall i=1,\dots, N$. A random orthogonal matrix requires that to either value $\varepsilon_i=\pm 1$ an equal probability of $P(\varepsilon_i=\pm 1)=1/2$ be associated. A rotator $R_{ij}(\Theta_{ij})$, itself an $N\times N$-orthogonal matrix, acts as a rotation in the $(i\!-\!j)$-$2$-plane solely, and is thus characterized by an angle $\Theta_{ij}$, while being restricted to the identity operator ${\mathbf{1}}_{2\times 2}$ on any of the other $(l\!\!-\!\!m)$-$2$-planes with either $l,m\neq i$ or $l,m\neq j$. The $\Theta_{ij}$ are independent random variables with a joint probability distribution proportional to~\cite{Anderson et al.},
\begin{equation}\label{angles}\nonumber
\prod_{j=2}^N\cos^{j-2}\Theta_{1j}\,\prod_{j=3}^N\cos^{j-3}\Theta_{2j}\dots\prod_{j=N}^N\cos^{j-N}\Theta_{N-1,j}\,,
\end{equation}
whereas the probability density of an angle $\Theta_{ij}$ is a {\textit{beta}} distribution, $\beta(x_{ij};\frac{a}{2},\frac{b}{2})$, with $\cos\Theta_{ij}={\sqrt{x_{ij}}}$, that is, $\beta(x_{ij};\frac{a}{2},\frac{b}{2})=x_{ij}^{a/2\,-1}(1-x_{ij})^{b/2\,-1}/B(\frac{a}{2},\frac{b}{2})$. As meant in the second line of (\ref{Eq:1}), these probability densities allow one to calculate averages over orthogonal matrices in a definite quantitative way, though somewhat probability density dependent. 
\par\medskip
For our purpose however, this unwanted probability density dependence will not affect our derivations as the full explicit form of the Haar measure is not required, but only the $D_\varepsilon$ matrix properties, and the {\textit{left-}} and {\textit{right- invariances}} of the Haar measure on $O_N(\mathbb{R})$ \cite{ {Anderson et al.}}. Denoting by $a^{ij}=a^{ij}(\dots,\Theta_{lm}\,,\dots)$ the matrix elements of (\ref{O}) as the reflection matrix $D_\varepsilon$ is omitted, one obtains,
\begin{equation}\label{11}
\bigl\langle{\cal{O}}^{ij}{\cal{T}}_j{\cal{O}}^{ik}{\cal{T}}_k\bigr\rangle_{\varepsilon,\Theta}=\bigl\langle\varepsilon_j a^{ij}(\Theta)\,\varepsilon_ka^{ik}(\Theta){\cal{T}}_j{\cal{T}}_k\bigr\rangle_{\varepsilon,\Theta}={\delta_{jk}}\bigl\langle a^{ij}(\Theta)a^{ik}(\Theta)\bigr\rangle_\Theta{\cal{T}}_j{\cal{T}}_k\,,
\end{equation}where, in the last equality, the average over the product of reflections $\varepsilon_j\varepsilon_k$ has been taken. That is, $<{\sqrt{z_i}}>$ is given by
\begin{equation}\label{12}
\sqrt{\lambda}\sum_{j,k=1}^N\bigl\langle{\cal{O}}^{ij}{\cal{T}}_j{\cal{O}}^{ik}{\cal{T}}_k\bigr\rangle_{\varepsilon,\Theta}= \sqrt{\lambda}\,\bigl\langle\sum_{j=1}^Na^2_{ij}(\dots\Theta_{lm}\dots)\bigr\rangle_\Theta {\cal{T}}_j^2 =\frac{\sqrt{\lambda}}{N}\,DC_{2f}\, {\mathbf{1}}_{3\times 3}\,,
\end{equation}where $C_{2f}$ stands for the quadratic Casimir operator eigenvalue on the fundamental representation, $C_{2f}=C_F=4/3$, and where $ {\mathbf{1}}_{3\times 3}$ is the identity matrix of format $3\times 3$. The last equality of (\ref{12}), independent of the index $i$, is a consequence  of the left- and right- invariances of the Haar measure on $O_N(\mathbb{R})$, as it is shown in Appendix C .

\par\medskip
In the same way, for the sub-leading piece of (\ref{zexpand}), one obtains, for all $i=1,\dots, N$,
\begin{equation}\label{13}
<z_i>\,=(\frac{\sqrt{\lambda}}{N})^2\left(\,(DC_{2f})^2+(DC_{3f})\right){\mathbf{1}}_{3\times 3}\,,
\end{equation} in which one notices the occurrence of the cubic Casimir operator eigenvalue in the fundamental representation of $SU_c(3)$,
\begin{equation}\label{14}
\sum_{a,b,c=1}^{N_c^2-1} d_{abc}\,t^at^bt^c\equiv C_{3f}{\mathbf{1}}_{3\times 3}\ .
\end{equation}The fully symmetric constants $d_{abc}$ are defined in the standard fashion, that is, in the case of interest, for $N_c=3$, $\{t_a,t_b\}=d_{abc}t_c+\frac{1}{3}\delta_{ab}$. Far less popular than $C_{2}$, the trilinear Casimir operator eigenvalue over a representation space specified by the {\textit{Young Tableaux}} parameters $(p,q)$ is given by  
\begin{equation}\label{15}C_{3}(p,q)=\frac{1}{18}(p-q)(2p+q+3)(2q+p+3),\end{equation}
and $C_{3f}=C_3(1,0)=10/9$ in the $SU_c(3)$ fundamental representation~\cite{Anirban}, whereas it is $C_{3a}=C_3(1,1)=0$, in the adjoint representation, $(1,1)$, another salient feature of distinction between QCD and the pure Yang-Mills case.
\par\medskip
At next to sub-leading order, $z_i\sqrt{z_i}$, corresponding to the $O_N(\mathbb{R})$-averaged value of $({\cal{O}}^{ij}({\mathbf{p}}){\cal{T}}_j)^6$, calculations become more intricate. One finds,
\begin{eqnarray}\label{16}
&& <z_i\sqrt{z_i}>\,= (\frac{\sqrt{\lambda}}{N})^3\,\biggl\lbrace\left(\,2(DC_{2f})^2+(DC_{2f})(DC_{3f})+\frac{4}{3}(DC_{3f})\right){\mathbf{1}}_{3\times 3}\nonumber\\ && +\sum_{k,j,l,h,m}\, d_{kjm}d_{khl}\, (T_jT_mT_hT_l+2 T_jT_hT_lT_m)\biggr\rbrace\,.
\end{eqnarray}While independent of $i$, the two last terms look somewhat puzzling as they seem to compromise the general structure of these dependences, in terms of $SU_c(3)$ algebraic invariants. To proceed, one may rely on the standard textbook values of the $d_{abc}$ coefficients~\cite{Yndurain}, and work out the following two identities,
\begin{equation}\label{17}\sum_{k,j=1}^8\, d_{kjj}=0\,,\ \ \ \ \ \ \ 
\sum_{j,m=1}^8\, d_{k'jm}d_{kjm}=\frac{5}{3}\,\delta_{k'k}\,,
\end{equation}of which, the second one can also be found in Ref.\cite{Close}. It is then easy to prove that (\ref{16}) indeed reduces to,
\begin{equation}\label{18}
<z_i\sqrt{z_i}>\,=(\frac{\sqrt{\lambda}}{N})^3\left(\,[2+(\frac{5}{6})^2](DC_{2f})^2+(DC_{2f})(DC_{3f})+{3}(DC_{3f})\right){\mathbf{1}}_{3\times 3}\,.
\end{equation}An analysis of the higher order terms of (\ref{new1}) is likely to become much more complicated and not very insightful either, as it is now made clear enough that $C_{3f}$-dependences will enter any $2n$-point non-perturbative fermionic Green's function and related amplitude, and this is the point of the present paper.

\section{\label{SEC:4} Some numerical insights}

Concerning numerical orders of magnitude, the result of (\ref{13}) shows that at sub-leading  order $({\sqrt{\lambda}/N})^2$, the trilinear Casimir operator $C_{3f}$ enhances the pure $C_{2f}$ contribution an amount of $ 15.6\%$,  to be compared to the $15\%$ at most, advocated in \cite{Cooper}, whereas at sub-sub-leading order $({\sqrt{\lambda}/N})^3$, $C_{2f}$ and $C_{3f}$ contributions to (\ref{18}) are identical to within $0.2\%$. \smallskip

Now, even at a very large absolute value of the $\lambda$-parameter of (\ref{z_i}), (see Appendix E, Eq.(\ref{range})), the contribution brought about by the $C_3$-dependence enhances the pure $C_{2f}$ contribution a small relative amount of,
\begin{equation}\label{deviation}
\frac{\sqrt{\lambda}}{N}\,\frac{DC_{3f}}{DC_{2f}+(\frac{\sqrt{\lambda}}{N})\,(DC_{2f})^2}\simeq 0.01\%\,,
\end{equation}where ${\sqrt{\lambda}/N}\simeq 0.012\%$, a value obtained at a strong coupling of $g=15$, $\mu^2/{\widehat{s}}\exp-(\mu b)^{2-\xi}\simeq~1$, $m=5MeV$ and ${\widehat{s}}=100 MeV$. Likewise, as compared to a pure linear dependence in $C_{2f}$ alone, one gets a less than $0.1\%$ relative deviation,
\begin{equation}\label{departure}
\frac{\sqrt{\lambda}}{N}\,\,\frac{(DC_{2f})^2+DC_{3f}}{DC_{2f}}\simeq 0.075\%\,.
\end{equation}

Though it is here dealt with {\textit{dynamical}} rather than {\textit{static}} quarks, so small departures from a pure $C_{2f}$-dependence comply with the roughly measured linear $C_{2f}$-dependence of \cite{Bali} (2nd paper),  as well as it supports the experimental $5\%$ of maximal deviation from the so-called {\textit{$C_{2f}$-scaling hypothesis}} advocated in \cite{Bali} (1st paper). 
\par\smallskip
The point however, is to estimate how the elementary relative deviations (\ref{deviation}) or (\ref{departure}) translate at the level of a whole monomial, and, further, to the full sum of them. This is quite long a numerical affair, two hints of which may be given here.
\par\medskip - Neglecting the sub-sub-leading correction (\ref{18}) to the leading and sub-leading ones, (\ref{11}) and (\ref{12}) respectively, and considering a monomial like $\xi_1^{q_1}\xi_2^{q_2}\cdots\xi_N^{q_N}$ associated to a distribution of the powers $N-1, N-2,\cdots, 1, 0$ for the $q_is$, one gets for (\ref{zexpand}) the expansion,
 \begin{eqnarray}\label{zexpand'}&&
+  (-{4\pi^2 m^2\over E^2})^N\biggl\langle\prod_{i=1}^N\, [1-i(-1)^{q_i}]{\sqrt{\pi}}\,\Gamma(\frac{2q_i+1}{4})\nonumber\\ &&\times\left(\,1+ \frac{-2\Gamma(\frac{2q_i-1}{4})}{\Gamma(\frac{2q_i+1}{4})}\,{\sqrt{z}}+\frac{8}{2q_i-3}\,z +{\cal{O}}(z^{\frac{3}{2}})\right)\biggr\rangle_{\!O_N(\mathbb{R})}\,,
\end{eqnarray}where (\ref{Anumbers}). (\ref{Onumbers}) as well as the relation $\Gamma(-\frac{1}{2})/\Gamma(\frac{1}{2})=-2$ with $\Gamma(\frac{1}{2})={\sqrt{\pi}}$ have been used. Introducing the shorthand notations,
\begin{equation}\label{shorthand}
a_i\equiv \frac{-2\Gamma(\frac{2q_i-1}{4})}{\Gamma(\frac{2q_i+1}{4})}\,,\ \ \ \ \ b_i\equiv \frac{8}{2q_i-3}\,,
\end{equation}equation(\ref{departure}) can be reported to the level of a whole monomial, giving, as a relative deviation $\delta_{C}$, to a pure $C_{2f}$-linear Casimir behavior,
\begin{equation}\label{deltaC3}
\delta_{C}=\frac{\sqrt{\lambda}}{N}\,\,(\frac{C_{3f}}{C_{2f}}+DC_{2f})\ \,\frac{\sum_1^Nb_i+\sum_2^Na_i\sum_1^{i-1}a_j}{\sum_1^Na_i}\,+ {\cal{O}}(\,(\frac{\sqrt{\lambda}}{N})^2)\,,
\end{equation}For the monomial considered, involving the powers $N-1, N-2,\cdots, 1, 0$, a long calculation yields a relative departure of,
\begin{equation}
\delta_{C}(N-1,N-2,\cdots, 1,0)\simeq -1.0\%\,,
\end{equation}which isn't that small a number if we keep in mind that there is an enormous number of monomials. However, a large number of monomials involve the same values for the $q_is$, attributed to different eigenvalues $\xi_is$ to be integrated upon. Alternate in $+$ and $-$ signs, these contributions to the full sum of monomials cancel out. This is made explicit within the following example, worked out in the easiest case of $N=4$.
\par\medskip
- At $N=4$, one has $2^{6}=64$ monomials, that reduce to $48$ due to a trivial cancellation. Each monomial $\pm \xi_1^{q_1}\xi_2^{q_2}\cdots\xi_{N-1}^{q_{N-1}}\xi_N^{q_N}$ is affected with a plus or a minus sign and these signs come about in an equal number of occurrences. Once integrated over the full spectrum of eigenvalues, the $\xi_is$, and the auxiliary field variables, (\ref{Meijer}) and (\ref{A11}), the contribution of a given monomial can be represented by a `word',
\begin{equation}\label{monom}
\pm \xi_1^{q_1}\xi_2^{q_2}\cdots\xi_{N-1}^{q_{N-1}}\xi_N^{q_N}\ \longrightarrow\  \pm(q_1q_2\cdots q_N)\,,\end{equation}with the sum of the $q_i's$ adding up to $N(N-1)/2=6$, in this case.
 In this way, the sum of all $48$ fully  integrated monomials admits the following symbolic representation,
\begin{eqnarray}\label{monomials'}&&+(3210)-(3201)+(3102)-(3120)+(3021)-(3012)-(2211)+(2202)  \nonumber \\ &&-(2103)+(2121)-(2022)+(2013)-(2220)+(2202)-(2112)+(2130)\nonumber \\ &&  -(2031)+(2022)+(1221)-(1212)+(1113)-(1131)+(1032)-(1023) \nonumber \\ &&-(2310)+(2301)-(2202)+(2220)-(2121)+(2112)+(1311)-(1302)\nonumber \\ && +(1203)-(1221)+(1122)-(1113)+(1230)-(1221)+(1122)-(1140)\nonumber \\ &&     +(1041)-(1032)-(0231)+(0222)-(0123)+(0141)-(0042)+(0033)\,.
\end{eqnarray}Averaged over $O_N(\mathbb{R})$, the sum of contributions (\ref{monomials'}) yields (\ref{new1}) taken at $N=4$, and it can be further expanded along (\ref{zexpand'}) in view of the smallnessss of ${\sqrt{\lambda}}/N\simeq 0.1\%$ at $N=4$. Now, clearly, the order along which the numbers $q_is$ appear in a `word' is irrelevant to the result: Contributions such as $(3210)$, $(1203)$, $(2031)$, for example, are one and the same contribution. Accordingly, a lot of cancellations take place and (\ref{monomials'}) eventually reduces to,
\begin{equation}\label{final'}2\times(0222)-2\times(0123)-(1212)+(0141)-(0042)+(0033)\,.\end{equation}
\par
As compared to a purely linear behavior of the fermionic 4-point function in $C_{2f}$ alone, the extra contributions in $C^2_{2f}$  and $C_{3f}$ are therefore in a ratio of,
\begin{equation}
- \frac{1798,211444}{147,543465}\ \frac{\sqrt{\lambda}}{4}\,\frac{\,DC_{3f}+(DC_{2f})^2}{DC_{2f}}=-7.5\%\,.\end{equation}
\par
Such a result doesn't comply exactly with the allowed experimental boundaries of $5\%$. However, several points must be taken into account: First, it holds at $N=4$, away from the realistic value of $N=32$. In a second place, it is obtained within the quenching and eikonal approximation. Thirdly, it is obtained on the basis of $\mu^2/{\widehat{s}}\exp-(\mu b)^{2-\xi}\simeq1$, a phenomenological input which is here deliberatly overestimated, given that $\mu$ the effective locality scale is not known at present. In effect, non-perturbative physics should take place at a distance $b\geq 1/\Lambda_{QCD}$, so that the exponential of $\exp-(\mu b)^{2-\xi}$ could reduce the result in a significant way if the effective locality scale is larger than $\Lambda_{QCD}$ (for example, such would be the case, would the effective locality scale $\mu$ be close to some chiral symmetry breaking estimates, a possibility under consideration \cite{tg'}).
\par
Clearly, a way has to be found to cope with the realistic value of $N=32$; work in this direction is in progress where, to begin with, the value of $N=32$ has already been reduced to the smaller value of $N=16$.

\section{\label{SEC:5} Conclusion}
The newly discovered property of effective locality is here explored in one of its consequences, namely, the color algebraic structure of the fermionic strong coupling Green's functions and amplitudes thereof. At variance with a whole series of non-perturbative approaches, the latter are found to display dependences on the full color algebraic content of the $SU_c(3)$-algebra, which is of rank 2. That is, not only $C_{2f}$, the quadratic Casimir operator in the fundamental representation of $SU_c(3)$ is involved, but also the trilinear one $C_{3f}$. In the absence of a {\textit{superselection rule}} that would prevent $C_{3f}$ to show up, one may think that in a way or another, such a result is natural \cite{{BMuller}, {AWipf}}. 
\par\medskip

At the exception of a few previous results \cite{{Nieuwenhuizen}, {Cooper}}, where similar quadratic and trilinear color-invariants were put forward as relevant to the description of non-perturbative calculations, this is quite unexpected a result and this is why it matters to present some details of its derivation.
\par
This is the purpose of the current article where Appendices are used to keep a main text from being overwhelmed with too many technical intricacies. In Appendix A, for example, one summarizes the context in which the starting expression (\ref{Eq:1}) of the current analysis is derived, whereas in Appendix F, one can realize that the welcome structure of strong coupling fermionic Green's functions, in terms of finite sums of finite products of Meijer functions is a non-trivial result: It is induced by $O_N(\mathbb{R})$-averaging. Over exponentials of non commuting $SU_c(3)$ generators, $O_N(\mathbb{R})$-averaging operates in exactly the same formal way as {\textit{Time-ordering}} does on exponentials of non-commuting field operators. We think that these structures shed some new lights on the genuine nature of effective locality, as should be explained elsewhere.
\par\medskip
As it comes to numbers, our preliminary results seem to fall into the experimentally allowed range of possible deviations from a sole quadratic Casimir dependence of non-perturbative fermonic Green's functions. Cubic Casimir contributions appear only as subleading effects. On a purely theoretical point of view though, a comparison of Equations (\ref{12}) and (\ref{13}) shows that a pure linear $C_{2f}$-dependence is but an approximation, however excellent. These results are derived within two major approximations that are eikonal and quenching. While the former one is appropriate to describe high energy scattering processes, the latter doesn't preserve unitarity. 
\par
First insights \cite{prep}, obtained by relaxing the quenching approximation display the kind of compensation mechanisms that are peculiar to the unitarity constraint. But, even though modified, $C_{3f}$-dependences appear to persist under unitarity restoration. Of course, a more quantitative determination of these dependences is in order, and will be addressed in a forthcoming publication \cite{prep}. Though subleading, it is quite likely that $C_3$ dependences be the hallmarks of genuine non-perturbative calculations.

\begin{acknowledgments}
It is a pleasure to thank B. M\"uller and A. Wipf for interesting discussions, and M. Gattobigio for providing valuable references.
% put your acknowledgments here.
\end{acknowledgments}

\appendix
\section{\label{AppA}Context of Equation (\ref{Eq:1})}
\par
Equation (\ref{Eq:1}) is non-trivial and is the result of lengthy developments, whose guiding lines will be recalled briefly. Relying on standard functional methods, identities and notations, the QCD generating functional can be brought to the explicit form of \cite{QCD1},
\begin{eqnarray}\label{Z}
\mathfrak{Z}_{\mathrm{QCD}}[j,\eta, {\bar{\eta}}] = \mathcal{N} e^{\frac{i}{2} \int{j \cdot \mathbf{D}_{c}^{(\zeta)} \cdot j}}  \int{\mathrm{d}[\chi] \, e^{ \frac{i}{4} \int{ \chi^{2} }} } \, \left. e^{\mathfrak{D}_{A}^{(\zeta)}} \cdot e^{-\frac{i}{2} \int{\chi \cdot \mathbf{F} + \frac{i}{2} \int{ A \cdot \left( -\partial^{2}\right) \cdot A} }} \cdot e^{i\int{\bar{\eta} \cdot \mathbf{G}_{\mathrm{F}}[A] \cdot \eta} + \mathbf{L}[A]}\right|_{A = \int{\mathbf{D}_{\mathrm{F}}^{(\zeta)} \cdot j} }
\end{eqnarray}where $\mathfrak{D}_{A}^{(\zeta)} = - \frac{i}{2} \int{\frac{\delta}{\delta A} \cdot \mathbf{D}_{\mathrm{F}}^{(\zeta)} \cdot \frac{\delta}{\delta A}}$ is the functional {\textit{linkage operator}}, and $\mathbf{D}_{\mathrm{F}}^{(\zeta)}$ the gluonic Feynman propagator in a given covariant  gauge with parameter $\zeta$. The form (\ref{Z}) may look a bit unusual as, though equivalent, functional integration is much more customary than functional differentiation. Deriving (\ref{Z}), $\chi^{a}_{\mu\nu}(x)$-Halpern fields have been introduced to {\textit{linearize}} the original field-strength part of the QCD lagrangian as in \cite{Halpern1977a,Halpern1977b,{Reinhardt}},
\begin{equation}\label{Halpern}
e^{-\frac{i}{4} \int{\mathbf{F}^{2}}} = \mathcal{N'} \, \int{\mathrm{d}[\chi] \, e^{ \frac{i}{4} \int{ \left(\chi_{\mu \nu}^{a}\right)^{2} + \frac{i}{2} \int{ \chi^{\mu \nu}_{a} \mathbf{F}_{\mu \nu}^{a}} } } }\,.
\end{equation}
The approximation of {\textit{quenching}} amounts to take to $0$ the fermonic determinant  functional ${\mathbf{L}[A]}$ of (\ref{Z}), and redefine the normalization accordingly. Fermonic $2n$-point Green's functions are obtained in the usual way, by differentiating with respect to the fermonic sources $\eta$, ${\bar{\eta}}$, and then by cancelling the sources, $\eta={\bar{\eta}}=j=0$. In (\ref{Z}), this operation brings down functionals $\mathbf{G}_{\mathrm{F}}(x_i,y_i|A)$ which admit exact {\textit{Schwinger-Fradkin representations}}, \cite{{QCD1},{QCD-II},{QCD5}, {QCD6},{QCD5'}}. One has for instance of a mixed representation (space-time and momentum),\begin{eqnarray}
& & {\langle p| \mathbf{G}_{F}[A] |y \rangle}\label{Fradkin}  = e^{-i p \cdot y} \cdot i
\int_{0}^{\infty}{ds \ e^{-is m^{2}}} \cdot e^{- \frac{1}{2} \Tr{\ln{\left(2h\right)}} }\\
\nonumber && \quad \times \int{d[u]} {\left\{ m - i \gamma \cdot \left[
p - g A(y-u(s)) \right] \right\}} \cdot e^{\frac{i}{4} \int_{0}^{s}{ds' \, [u'(s')]^{2} } } \cdot e^{i p \cdot u(s)} \\ \nonumber & & \quad \times  \framebox{$\left( e^{g \int_{0}^{s}{ds' \sigma \cdot F(y-u(s'))}} \cdot e^{-ig \int_{0}^{s}{ds' \, u'(s') \cdot A(y-u(s'))}} \right)_{+} $}
\end{eqnarray}where $s$ is the {\textit{Schwinger proper time}},  $u(s)$, the Fradkin $4$-vector field and $h$ the function $h(s_1,s_2)=\int_0^s {\rm{d}}s'\,\Theta(s_1-s')\Theta(s_2-s')$. The interesting point is in the last line where only quadratic dependences on the gluonic field $A^a_\mu(x)$, at most, appear, but in an $s$-ordered exponential, as indicated by the prescription $+$. \par
 The eikonal approximation, then, takes (\ref{Fradkin}) to a simpler form with, to wit, such an $s$-ordered exponential as,
\begin{equation*}\exp\left(ig\,p^\mu\!\int_0^s {\rm{d}}s'\,A^a_\mu(y-s'p)\,T^a\right)_+\,,
\end{equation*}where the $T^as$ stand for the $SU(3)$-Lie algebra generators in a suitable representation. By introducing 2 subsidiary functional integrations, on $\alpha^a(s)$ and $\Omega^a(s)$ fields, say, it is possible to take the $A^a_\mu(x)$-field dependences out of the $s$-ordered exponential, 
\begin{equation}\label{ordered}
\left(e^{ig\,p^\mu\!\int_{-\infty}^{+\infty} {\rm{d}}s\,A^a_\mu(y-sp)\,T^a}\right)_+=\mathcal{N}\int {\rm{d}}[\alpha]\int{\rm{d}}[\Omega]\, e^{-i\!\int_{-\infty}^{+\infty} {\rm{d}}s\,\Omega^a(s)[\alpha^a(s)-gp^\mu A^a_\mu(y-sp)]}\left(e^{i\int_{-\infty}^{+\infty}{\rm{d}}s\alpha^a(s)T^a}\right)_+
\end{equation}and to {\textit{complete the quadrature}} in an exact way (i.e., to get the full result of the linkage operations that are involved in (\ref{Z})). Now, to be guaranteed to deal with the proper representation (\ref{Fradkin}) or its eikonal approximated form (\ref{ordered}), it is essential to perform the integrations on subsidiary $\alpha^a(s)$ and $\Omega^a(s)$ variables in an exact way: This is one of the most favourable circumstances which, a few steps later, are met in these calculations \cite{QCD6}. 
\par\smallskip
Then, the $A^a_\mu(x)$-field quadrature yields the expression, 
\begin{equation}\label{quadrature}
e^{-\frac{i}{2g}\int {\rm{d}}^4z\, \mathcal{Q}(z)\,\left(f\cdot\chi(z)\right)^{-1}\mathcal{Q}(z)}\,e^{-\frac{1}{2}Tr\,[-g(f\cdot\chi)\mathbf{D}_{\mathrm{F}}^{(\zeta)}]}\,,\ \ \ \ \ \left(f\cdot\chi\right)^{ab}_{\mu\nu}=f^{abc}\chi^c_{\mu\nu}\,,
\end{equation}where, in the case of a fermonic $4$-point Green's function, 
 \begin{equation}\label{currents}
 \mathcal{Q}^a_\mu=-\partial^\nu\chi^a_{\mu\nu}+g[\mathcal{R}^a_{1\mu}+\mathcal{R}^a_{2\mu}]\,.\end{equation}In (\ref{currents}), the $\mathcal{R}^a_{i\mu}$, $i=1,2$, stand for the eikonal representation involving (\ref{ordered}), as the $A^a_\mu(x)$-field quadrature has been performed. \par\smallskip
 In the pure Yang Mills case, a form such as (\ref{quadrature}) and (\ref{currents}) has been put forth in Ref.\cite{Reinhardt}.  In Refs.\cite{QCD6} and \cite{tg}, as well as in the current paper, the strong coupling limit of $g>>1$ is introduced in a somewhat `academic way', so as to get rid of the $\nabla\cdot\chi\, [f \cdot \chi]^{-1}\cdot\nabla\cdot\chi $-term of (\ref{quadrature}), that so far, we have not been able to treat on the same footing as the other terms. For the reason given just after Eq.(\ref{15}), though, one may argue that this pure gluonic term does not impinge on our results.
 
\par
Note that the $A^a_\mu(x)$-quadrature given above displays the property of effective locality, as the generated interactions of the $\mathcal{R}^a_{i\mu}s$ are local, mediated by the structure $(f\cdot\chi(z))^{-1}$, taken at the same space-time point $z$ \cite{QCD1}. The result of the quadrature (besides a residual dependence on $\mathbf{D}_{\mathrm{F}}^{(\zeta)}$ to be absorbed into the normalization constant), does no longer depend on the initial covariant gauge-field propagator: The same independence is obtained in a similar way with any other bare gauge-field propagator, in any gauge \cite{QCD-II}. Now, it is of utmost importance to emphasize that a series of functional differential identities allows one to prove that the same property of effective locality is satisfied also by the full non-approximated QCD theory \cite{QCD-II}. This property comes along with a mass scale $\mu$, (\ref{phi}), (the consequence of a theorem \cite{QCD6}) whose ultimate identity is still under investigation. 
 \par\smallskip\noindent
 In the $4$-point function case which illustrates the purpose of the current article, and at this level of approximation, the integrations that are left are those on proper-times $s_1$ and $s_2$ (made trivial by the theorem of \cite{QCD6}), on the subsidiary field variables $\alpha^a(s)$ and $\Omega^a(s)$, and on the $\chi^a_{\mu\nu}$-fields. 
 \par\smallskip
 Concerning the latter, it is convenient to define $iM$, the product,
 \begin{equation}\label{M}
 \sum_{a=1}^{N_c^2-1}{{\chi^a_{\mu\nu}}\otimes T^a} = iM\, ,\  \quad M_{ij}= M_{ji}\in \mathbb{R}\, ,\ \ 1\leq i,j\leq N\, ,\quad \tr M=0.
\end{equation}where $N=D\times(N_c^2-1)$ \cite{{Halpern1977a},{Halpern1977b}}, so that one has now to deal with an integration over {\textit{random matrices}} with measure \cite{{QCD6},{Mehta1967}},
\begin{eqnarray}\label{measure}
& & {\rm{d}}(\sum_{a=1}^{N_c^2-1}{\chi^a}_{\mu \nu}\otimes T^a) \nonumber \\ &=& i{\rm{d}}M= i{\rm{d}}M_{11}\,{\rm{d}}M_{12} \cdots {\rm{d}}M_{NN} \nonumber \\ &=&\nonumber i\left|\frac{ \partial(M_{11}, \cdots, M_{N\!N})}{\partial(\xi_1, \cdots, \xi_N, p_1, \cdots, p_{N(N-1)/2})}\right| \, {\rm{d}}\xi_1 \cdots {\rm{d}}\xi_N \, {\rm{d}}p_1 \cdots {\rm{d}}p_{N(N-1)/2} \\  &=& i\prod_{i=1}^{N} {\rm{d}}\xi_i  \prod_{i<j} |\xi_i-\xi_j|^{\kappa}\   {\rm{d}}p_1\  ..\ {\rm{d}}p_{N(N-1)/2}\, f(p)\,,
\end{eqnarray}where the $\xi_i s$ are the eigenvalues of $M$, while the $p_js$ complete the parametrization of a given matrix $M$ (that is, the original integration on matrices $M$ splits into an integration on its spectrum of eigenvalues and on the orthogonal group ${\!O_N(\mathbb{R})}$). In (\ref{measure}), one has a {\textit{Vandermonde determinant}} of (at $\kappa=1$, see Appendix B),
\begin{eqnarray}\label{VdM}
{\cal{P}}(\xi_1,\dots,\xi_{N})=\prod_{1\leq i<j\leq N} |\,\xi_i-\xi_j|=|\!\!\sum_{\mathrm{monomials}}\,\pm\ \xi_1^{q_1}\dots \xi_N^{q_N}\ | \ .
\end{eqnarray} The sum in the right hand side of (\ref{VdM}) comprises $2^{N(N-1)/2}$ terms. Each term, a monomial, is characterized (not in a unique way) by a given distribution of $q_i$-powers whose sum satisfies the constraint of an equal global degree of $N(N-1)/2$. Up to coefficients that are $q_i$-dependent, (\ref{Anumbers}), (\ref{Onumbers}), the monomials share the same algebraic color structure, which is that of the net Green's function and related amplitude. \smallskip
It is by integrating on a given eigenvalue $\xi_i$, that Meijer functions do appear according to \cite{pomme},
\begin{eqnarray}\label{Meijer}
\int_0^\infty\ {{\rm{d}}\xi_i}\ \xi_i^p\ e^{-\xi_i^2-{b_i\over \xi_i}}={\frac{1}{2 \sqrt{\pi}}}\, G^{30}_{03}\left(\frac{b_i^2}{4} \biggr| {p+1\over 2}, \frac{1}{2},0\right)\,.
\end{eqnarray}In (\ref{Meijer}), equality holds only for $p>0$ and $b>0$, whereas the right hand side is analytic both in the argument and the parameters of the Meijer function: This is the second most favorable circumstance of these calculations.
\par\medskip
A third most favorable circumstance comes about as one deals with integrations on the subsidiary field variables $\alpha^a(s)$ and $\Omega^a(s)$ that have to be carried out in an exact way, as quoted above. Defining $V'_i={\cal{O}}V_i$, where ${\cal{O}}$ is an $N\times N$ orthogonal matrix, and where $V_i=p_i\otimes\Omega^a_i(0)$ with the index $i=1,2$ labelling the 2 scattering quarks, then the Meijer function argument in (\ref{Meijer}), $b^2_i/4$, turns out to be proportional to $[(V'_1)^i(V'_2)^i]^2$.
\par\bigskip
Now, in view of (\ref{ordered}) or expression (\ref{E1}), and dropping the $'$-notations for short, these dependences on the $V^is$, must be folded into the following two integrations,
\begin{equation}\label{A11}
\int{\mathrm{d} V^i_1}\, e^{-i{{{\hat{\alpha}}^i_1\, V^i_1}\over E-p}} \int{\mathrm{d} V^i_{2} \, e^{-i{{{\hat{\alpha}}^i_2\, V^i_2}\over E+p}} }{\frac{1}{2 \sqrt{\pi}}}\, G^{30}_{03}\left(\frac{b_i^2}{4} \biggr| {2q_i+1\over 4}, \frac{1}{2},0\right)\,,\end{equation} whose exact results are provided by the Tables \cite{{Erdelyi1953},{Erdelyi1954}}. Once all pieces are glued togeher, the result is given by (\ref{Eq:1}) in the main text, the expression the current article is starting from.

\section{\label{AppB}Absolute values in the Vandermonde determinant}
\par
Passing from the functional integrations over the Halpern fields $\chi^a_{\mu\nu}$ to a summation over an algebra of  random matrices, one encounters the {\textit{Jacobian}} (\ref{measure}),\cite{QCD6,Mehta1967}, carrying the Vandermonde determinant (\ref{VdM}),
\begin{equation}\label{A2'}
{\cal{P}}(\xi_1,\dots,\xi_{N})=\prod_{1\leq i<j\leq N} |\,\xi_i-\xi_j|^{\kappa}\,.
\end{equation}
Now, contrarily to hermitian or symplectic matrices for which the parameter values are $\kappa=2$ {{and}} $\kappa=4$ respectively, for a real symmetric matrix, we have $\kappa=1$, and analyticity seems to be lost. But that is not the case. Since matrices $M$ are real traceless symmetric, their {\textit{spectra}} can be written in the following way,
\begin{equation}\label{spectra}
{\mathrm{Sp}}\,M=\bigl\lbrace\,(\xi_i,\xi_{N-i+1}=-\xi_i)\bigr\rbrace\ , \forall i=1,2,\dots, N/2\,.
\end{equation}That is, under the form of $N/2$ pairs of equal and opposite eigenvalues, the eigenvalue zero being degenerate, with a mutiplicity greater or equal to two. As a consequence, the Vandermonde determinant (\ref{A2'}) boils down to a form,
\begin{eqnarray}\label{}
{\cal{P}}(\xi_1,\dots,\xi_{N})=\prod_{1\leq i<j}^N |\xi_i-\xi_j|^\kappa =(\prod_{i=1}^{N/2}2\xi_i)^\kappa\, (\prod_{1\leq i<j}^{N/2} (\xi^2_i-\xi^2_j)^2\,)^\kappa\,,
\end{eqnarray}which, even at $\kappa=1$, is obviously analytic. Unfortunately, however, the symmetry properties of the {\textit{spectra}} (\ref{spectra}) are broken by interaction terms in such a way that the exact integration {\textit{formulae}} of Ref.\cite{Erdelyi1954} can no longer be used. In order to show that the absolute value prescription of (\ref{A2'}) can be dropped without prejudice for our concern, one must accordingly construct another argument, more in the line of a `physicist proof'.
\par\medskip

Relying on a standard representation of the Heaviside step-function such as
\begin{equation}\label{A1}
\Theta(x)=\frac{1}{2i\pi}\int_{-\infty}^{+\infty}{\rm{d}}z\frac{e^{izx}}{z-i\varepsilon}
\end{equation}one can write the Vandermonde determinant (\ref{A2'})) as,
\begin{equation}\label{A2}
{\cal{P}}(\xi_1,\dots,\xi_{N})=\frac{1}{\pi}\sum_{k=0}^\infty\frac{(-1)^k}{(2k+1)!}\int_{-\infty}^{+\infty}{\rm{d}}z\frac{z^{2k+1}}{z-i\varepsilon}\prod_{1\leq i<j\leq N}( \xi_i-\xi_j)^{2k+2}\,.
\end{equation}Now, in the main text, (\ref{Eq:1}) is obtained by integration over the the whole spectrum of $\xi_i$-eigenvalues, with, instead of (\ref{A2'}), the sole term $\prod_{1\leq i<j\leq N}\,( \xi_i-\xi_j)$, subsequently expanded into as a sum of monomials,
\begin{equation}\label{A3}
\prod_{1\leq i<j\leq N}( \xi_i-\xi_j)=\sum_{\mathrm{monomials}}\pm\  \xi_1^{q_1}\xi_2^{q_2}\dots \xi_N^{q_N}\,.\end{equation}Bringing (\ref{A3}) to any power of $2k+2$, one gets,
\begin{equation}\label{A4}
\prod_{1\leq i<j\leq N}( \xi_i-\xi_j)^{2k+2}=\sum_{\mathrm{monomials}} {\cal{C}}_{q_1\{k\}\dots q_N\{k\}}\  \xi_1^{q_1\{k\}}\xi_2^{q_2\{k\}}\dots \xi_N^{q_N\{k\}}\,,\end{equation}where for every value of $k$, the coefficients ${\cal{C}}_{q_1\{k\}\dots q_N\{k\}}$ as well as the powers $q_i\{k\}$ can be written in closed, explicit form. Now, clearly, the coefficients ${\cal{C}}_{q_1\{k\}\dots q_N\{k\}}$ are irrelevant to the point under consideration, and for the powers $q_i\{k\}$, the only property of concern is that it preserves integrability \cite{Erdelyi1954}, that is, $\forall k$, one must have $\ q_i\{k\}\geq q_i$, which is obviously satisfied ($q_i\{k\},q_i\in\mathbb{N}$). 
\par\medskip
With this representation of the absolute value prescription, (\ref{Eq:1}) should therefore be replaced by,
\begin{eqnarray}\label{A5}
&& \frac{1}{\pi}\sum_{k=0}^\infty\frac{(-1)^k}{(2k+1)!}\int_{-\infty}^{+\infty}{\rm{d}}z\frac{z^{2k+1}}{z-i\varepsilon}\sum_{\mathrm{monomials}}\,{\cal{C}}_{q_1\{k\}\dots q_N\{k\}}\prod^{\sum q_i\{k\}=N(N-1)(k+1)}_{1\leq i\leq N}\ [1-i(-1)^{q_i\{k\}}] \nonumber\\  & & \!\!  \times\  C\int{\rm{d}}p_1\  ..\ {\rm{d}}p_{N(N-1)/2}\ f(p_1,\dots, p_{N(N-1)/2})\,\! \int_0^{+\infty} {\rm{d}}\alpha_1^i\ {\sin[\alpha_1^i({\cal{OT}})_i]\over \alpha_1^i}\int_0^{+\infty} {\rm{d}}\alpha_2^i\ {\sin[\alpha_2^i({\cal{OT}})_i]\over \alpha_2^i} \nonumber\\  & & \quad \times \, G^{23}_{34}\left( \left.{iN_c}\left({ \alpha_1^i\alpha_2^i \over g\varphi(b) }\right)^2{{\hat{s}}({\hat{s}}-4m^2)\over 2m^4} \right|
%\matrix{{(3-2q_i)/ 4},&{1/ 2},&1\cr {1/ 2},& {1/ 2},&1,&1\cr}
\begin{array}{cccc}
  \frac{3 - 2 q_i\{k\}}{4}, & \frac{1}{2}, & 1, &  \\
   \frac{1}{2}, & \frac{1}{2}, & 1, & 1
\end{array}
\right)\,.
\end{eqnarray}That is, even though a $k$- series has been introduced, as well as a subsidiary $z$-integration, the expression (\ref{A5}) already displays that all terms share the same algebraic structure which is that of the second line of both (\ref{Eq:1}) and (\ref{A5}). This structure can accordingly be analyzed on the simpler form (\ref{A3}). In full rigor, this conclusion would require that integrations on $\alpha_1^i$ and $\alpha_2^i$ be carried out, so as to control a possible change of dependences generated by the shift $q_i\rightarrow q_i\{k\}$: That it doesn't happen to be so is a by-product of Appendix E, Eq.(\ref{easy}): With $q_i\{k\}\geq q_i$, replacing $q_i$, the numbers $A^h_i$ and $O_{ih}$ of (\ref{Anumbers}) and (\ref{Onumbers}) undergo the replacements of  $A^h_i(q_i)\longrightarrow A^h_i(q_i\{k\}) $ and $O_{ih}(q_i)\longrightarrow O_{ih}(q_i\{k\}) $, leaving unaffected the (at least leading order) Casimir operator dependences resulting from the $O_N({\mathbb{R}})$-average of (\ref{zexpand}). Another proof will be given elsewhere \cite{prep}.
\par\bigskip

\section{\label{AppC} $O_N({\mathbb{R}})$-averages}
\par
As a generic example, one can consider $<~a^2_{ij}a^2_{ik}>_\Theta$, the $O_N({\mathbb{R}})$-averaged value of the product $a^2_{ij}({\mathbf{\Theta}})a^2_{ik}({\mathbf{\Theta}})$, where all 3 indices $i,j,k$ are fixed, and where ${\mathbf{\Theta}}$ stands for the collective dependences on the angles $\Theta_{lm}$, as explicited in (\ref{O}). Let $R_{lk}(\pm\pi/2)$ be the rotator acting in the $l-k$ plane solely, taking the unit vector $|k\!>$ to the unit vector $|l\!>$, through an appropriate rotation of angle $\Theta_{lk}=\pm \pi/2$; the restriction of $R_{lk}(\pm\pi/2)$ to all of the other possible $2$-planes is just the unit operator ${\mathbb{I}}_2$. Over the whole space spanned by the orthonormal basis $\{|i>\,,i=1,\dots,|N>\}$, the standard {\textit{resolution of unity}} holds, ${\mathbb{I}}_N=\sum_{i=1}^N|i><i|$. 
\par\medskip
Thus, recalling that ${\cal{O}}_{ij}=\varepsilon_j a_{ij}$, with $\varepsilon_j=\pm 1\,,\forall j=1,\dots,N$, one has
\begin{equation}\label{B1}
<~a^2_{ij}a^2_{ik}>=C\int_{ O_N({\mathbb{R}})}\ \mu_H({\cal{O}})\  {\cal{O}}^2_{ij}\,{\cal{O}}^2_{ik}\,,
\end{equation}where $\mu_H$ is a standard and shorthand notation for the Haar measure on $O_N({\mathbb{R}})$. Within customary Dirac's notations, (\ref{B1}) can be written as,
\begin{equation}\label{B2}
<~a^2_{ij}a^2_{ik}>=C\int_{ O_N({\mathbb{R}})}\ \mu_H({\cal{O}})\  (<i|{\cal{O}}|j>)^2\,(<i|{\cal{O}}|k>)^2\,.
\end{equation}Integration on $O_N({\mathbb{R}})$ can be right-shifted by multiplying any matrix ${\cal{O}}$ to the right, with another orthogonal matrix ${\cal{O'}}$; that is, ${\cal{O}}\rightarrow {\cal{O}} {\cal{O'}}$, 
\begin{equation}\label{B3}
<~a^2_{ij}a^2_{ik}>=C\int_{ O_N({\mathbb{R}})}\ \mu_H({\cal{O}}{\cal{O'}})\  (<i|{\cal{O}}{\cal{O'}}|j>)^2\,(<i|{\cal{O}}{\cal{O'}}|k>)^2\,.
\end{equation}By choosing ${\cal{O'}}=R_{lk}(\pm\pi/2)$, one obtains
\begin{eqnarray}\label{B4}
 &&<~a^2_{ij}a^2_{ik}>=C\int_{ O_N({\mathbb{R}})}\ \mu_H({\cal{O}})\  (\sum_{h=1}^N<i|{\cal{O}}|h><h|R_{lk}|j>)^2\,(\sum_{h'=1}^N<i|{\cal{O}}|h'><h'|R_{lk}|k>)^2\nonumber\\  &&\qquad\qquad\quad\!= C\int_{ O_N({\mathbb{R}})}\ \mu_H({\cal{O}})\  {\cal{O}}^2_{ij}\,{\cal{O}}^2_{il}\nonumber\\ &&\qquad\qquad\quad\! =<~a^2_{ij}a^2_{il}>
\end{eqnarray}where the right-invariance of the Haar measure is used in the first equality, the definition of the particular orthogonal matrix $R_{lk}$ is used in the second one, and (\ref{B1}) in the last equality. In the same way, relying on the left-invariance of the Haar measure, the distinguished left side index $i$ can be shifted to any other of its possible values in the set$\{1,2,\dots,N\}$, establishing the uniform probability distribution of these products~\cite{Anderson et al.}. \par 
From this, and keeping in mind that the $a_{ij}$ can be looked upon as the components of orthonormalized $N$-vectors, it is easy to show (up to some overall multiplicative factors of $\pi$ which are not relevant to our considerations) that  one has $<a_{ij}^2>=1/N$, $<a^2_{ij}a^2_{ik}>=1/N^2$, $<a^2_{ij}a^2_{ik}a^2_{il}>=1/N^3$, etc..

\section{\label{AppD}Generalization to $2n$-point fermionic Green's functions}
\par
In the case of $2n$-point fermionic Green's functions, one has $n$-fermionic propagators, that is $n$ Fradkin field variables $u_i(s_i)$, and likewise, $n$-subsidiary fields $\alpha_i(s_i)$ and $\Omega_i(s_i)$.
\par
Ignoring renormalization effects for the time being, one has accordingly $C_n^2=n(n-1)/2$ $2$ by $2$-interaction terms~\cite{QCD6},
\begin{eqnarray*}
&& \frac{i}{2} g \int{\mathrm{d}^{4}w \, \int_{0}^{\bar{s_k}}{\mathrm{d}s_{k} \,  \mathrm{u_k}'_{\mu}(s_{k})\int_{0}^{\bar{s_l}}{\mathrm{d}s_{l}  \, {\mathrm{u_l}}'_{\nu}(s_{l}) \,}}} \\ \nonumber && \quad \times\, \Omega_k^{a}(s_{k}) \, {\Omega}_l^{b}(s_{l}) \,  \left. (f \cdot \chi)^{-1}(w) \right|^{\mu\nu}_{ab} \\ \nonumber && \times \, \delta^{(4)}(w_{} - y_{k} + \mathrm{u_k}(s_{k})) \, \delta^{(4)}(w_{} - y_{l} + {\mathrm{u_l}}(s_{l})),
\end{eqnarray*}where $1\leq k<l\leq n$. As demonstrated in Ref.\cite{QCD6}, each of these $2$ by $2$-interactions select a given point of Minkowski spacetime, $w_{kl}$, say, where that effective and local $k-l$ interaction is taking place, mediated by the structure $(f \cdot \chi)^{-1}(w_{kl})$. Such Green's functions therefore are proportional to,
\begin{eqnarray*}
&& \prod_{1\leq k<l\leq n}\int{\mathrm{d}^{(N_c^2-1)} \alpha_k} \int{\mathrm{d}^{(N_c^2-1)}  \alpha_{l} } \int{\mathrm{d}^{(N_c^2-1)}\, \Omega_k} \int{\mathrm{d}^{(N_c^2-1)}\,  \Omega_{l} \, e^{-i\alpha_k\cdot\Omega_k}\, e^{-i\alpha_{l}\cdot\Omega_{l}}\, e^{i\alpha_k\cdot T}\ e^{i\alpha_{l}\cdot T} } \nonumber \\ && \times\!\!\!\prod^{1\leq a\leq N_c^2-1}_{ 0\leq\mu<\nu\leq 3}\int{\mathrm{d}[\chi^a_{\mu\nu}(w_{kl})] \, \det[gf\cdot\chi(w_{kl})]^{- \frac{1}{2}}\ e^{ \frac{i}{4}\,\chi^2(w_{kl})+ig\varphi(b_{kl})\ \Omega_k^a[f \cdot \chi(w_{kl})]^{-1}\vert^{ab}_{30}\ \Omega^b_{l}} },
\end{eqnarray*}where, following the same steps as taken in Ref.\cite{QCD6}, one has now for each pair of different indices $k$ and $l$ in the set $\{1,\dots,n\}$,
\begin{equation}\label{C3}
\varphi(b_{kl})=({\mu/{\sqrt{\hat{s}_{kl}}}})\,e^{-[\mu\, b_{kl}]^{2-\xi}}.
\end{equation}In the expression above, as a straight forward generalization of a $4$-point fermionic Green's function, one defines $\hat{s}_{kl}=(p_k+p_l)^2$, and $b_{kl}={y_k}_\perp-{y_l}_\perp$ is taken in the $k-l$ center of mass system. From here, following again the calculation of Ref.\cite{QCD6}, one arrives for a given monomial of the expanded Vandermonde determinant at a structure of,
\begin{eqnarray*}&& \prod_{1\leq k<l\leq n}(-16\pi{m^2\over E_{kl}^2})^N\prod_{i=1}^N\,[1-i(-1)^{q_i}]\ \biggl\langle\,\int_0^{+\infty} {\rm{d}}\alpha_k^i\,\int_0^{+\infty} {\rm{d}}\alpha_l^i\ {\sin[\alpha_k^i({\cal{OT}})_i]\over \alpha_k^i}\ {\sin[\alpha_l^i({\cal{OT}})_i]\over \alpha_l^i} \nonumber\\  & & \quad \times \, G^{23}_{34}\left( \left.{32N_c\over -i}\left({ \alpha_k^i\,\alpha_l^i \over g\varphi(b_{kl}) }\right)^2{E_{kl}^2\,p_{kl}^2\over m^4} \right|
%\matrix{{(3-2q_i)/ 4},&{1/ 2},&1\cr {1/ 2},& {1/ 2},&1,&1\cr}
\begin{array}{cccc}
  \frac{3 - 2 q_i}{4}, & \frac{1}{2}, & 1, &  \\
   \frac{1}{2}, & \frac{1}{2}, & 1, & 1
\end{array}
\right)\biggr\rangle,
\end{eqnarray*}where $|p_k|=|p_l|\equiv p_{kl}$, and $E_k=E_l\equiv E_{kl}$, in each of the $2$ to $2$-$`k-l'\!-$ center of mass system. A $2n$-point fermionic Green's function therefore is given as the product of $C_n^2$- $2$ by $2$-building blocks that, up to their own kinematical variables, are identical to that of the starting expression (\ref{Eq:1}) (as should appear clear in Appendix F, in order to be able to write the above generalization of (\ref{Eq:1}), the results of Appendix F have to be anticipated).

\section{\label{AppE}Integration of expression (\ref{Eq:1})}
\par
The same steps as followed in \cite{QCD6} can be taken here to bring the second and third lines of (\ref{Eq:1}) into the form,
\begin{eqnarray*}
&&\pm (-{16\pi m^2\over E^2})^N\ \biggl\langle\,\prod_{i=1}^N\, [1-i(-1)^{q_i}] \int_{0}^{\infty} {\rm{d}}\alpha_1^i\ {\sin\,[\kappa\,\alpha_1^i({\cal{OT}})_i]}\nonumber \\  &&\times\int_{0}^{\infty} {\rm{d}}\alpha_2^i\ {\sin\,[\kappa\,\alpha_2^i({\cal{OT}})_i]}\,G^{23}_{34} \left( (\alpha_1^i \alpha_2^i)^2 \biggr\vert
%\matrix{{0},&{(1-2q_j)/ 4}\cr {0},& {0},&{1/2}\cr}
\begin{array}{cccc}
  0, & \frac{1 - 2 q_i}{4}, & \frac{1}{2}    \\
  \frac{1}{2}, & \frac{1}{2}, &0, & 0
\end{array}
\right)\biggr\rangle,
\end{eqnarray*}
where $\kappa$ is here introduced as a shorthand for the constant $\sqrt{{{g\varphi{\sqrt{-i}}\over{\sqrt{8N_c}}}}{m^2\over Ep}}$. For the sine function's arguments, no summation is understood over $i$, which is a {\textit{distinguished}} index, and as usual, the overall brackets denote the $O_N({\mathbb{R}})$-average. Also, with respect to (\ref{Eq:1}), {\textit{formula}} 5.3.1(8) of Ref.\cite{Erdelyi1953} has been used. \par
In the end, parameters are such, in the {\textit{sine}} functions as well as in the Meijer special functions, the array of numbers $\{ \, a_r(i) \,  \}$ and $\{  \, b_s \,  \}$, that {\textit{formula}} ${{20.5.}}(7)$ of Ref.~\cite{Erdelyi1954} can be used twice so as to take the above expression into the form of,
\begin{eqnarray*}
&& \nonumber\pm  (-{16\pi^2 m^2\over E^2})^N\ \biggl\langle\,\prod_{i=1}^N\, [1-i(-1)^{q_i}]\ \left({{\sqrt{32iN_c}}\,Ep\over {m^2}}\right) \\ &&\times\, \ \frac{[({\cal{OT}})_i]^{-2}}{g\varphi(b)}\,G^{52}_{47}\!\left( [{  g\varphi(b)\over {\sqrt{512iN_c}}  }{m^2\over Ep}]^2[({\cal{OT}})_i]^{4}\, \biggr|
%\matrix{1,&{1},&{1/2}\cr 1,&1,&1, {(2q_j+3)/ 4},&1/2,&{1/2}\cr}
\begin{array}{ccccccc}
  1, & 1, &             & \frac{1}{2} ,    & \frac{1}{2}             &  \\
   \frac{1}{2}, &  \frac{1}{2}, &  \frac{1}{2},    & \frac{2 q_i +3}{4}, & 1, & 1,& 1
\end{array}
\!\right)\biggr\rangle
\end{eqnarray*}

\noindent where the inversion formula {{5.3.1}}(9) of Ref.~\cite{Erdelyi1953} has been used in order to express the result in a form suited to a $ |z|< 1$ expansion of the Meijer function. 
\par
Again, in the expression above, the array of numbers is such that {{5.3.1}}(7) of Ref.~\cite{Erdelyi1953} can be used twice so as to eventually take it to the much simpler form of,
\begin{eqnarray}\label{easy}
&& \nonumber\pm  (-{16\pi^2 m^2\over E^2})^N\ \biggl\langle\,\prod_{i=1}^N\, [1-i(-1)^{q_i}] \\ &&\times\, \left({{\sqrt{32iN_c}}\,Ep\over {m^2}}\right) \frac{[({\cal{OT}})_i]^{-2}}{g\varphi(b)}\,G^{30}_{03}\!\left( [{  g\varphi(b)\over {\sqrt{512iN_c}}  }{m^2\over Ep}]^2[({\cal{OT}})_i]^{4}\, \biggr|\frac{1}{2}, \frac{3+2q_i}{4},1\!\right)\biggr\rangle
\end{eqnarray}

 Note that the integration {\textit{formulae}} of Refs.\cite{{Erdelyi1954},{Erdelyi1953}} have been extended to matrix-valued arguments. That this makes sense results from an available expression of Meijer's special functions, suited to the case of $ |z_i|< 1$. In the two previous forms, the Meijer function argument, $z_i$, is proportional to a quantity denoted by $\lambda$ in the main text, and defined in (\ref{z_i}). Even at sufficiently large values of the coupling, $g= 10-20$, for instance,  this number is very small in the range of moderate sub-energies $\hat{s}$. In that single quark model, taking a light dynamical quark mass of $m\simeq 5$MeV, and a moderate sub-energy ${\sqrt{\hat{s}}}\simeq 100$ MeV, even at a (very) large non-perturbative mass scale of about $300\,MeV$ for $\mu$ (corresponding to a very large estimate of the chiral condensate, for example), one has 
 \begin{equation}\label{range}0.013\leq \lambda\leq 0.13\,,\end{equation} while the matrix elements of $[({\cal{OT}})_i]^{4}$ are themselves of order unity: For all $i,j$, in effect, one has $|{\cal{O}}^{ij}({\bf{\Theta}})|\leq 1$, and so are also all of the matrix element of the $T_j$, for all $j=1,\dots,N$; and thus, {\textit{a fortiori}}, their product.

\par\medskip
For $|z|<1$, a Meijer function such as $G_{03}^{30}$ can be written as \cite{{Erdelyi1953}},
\begin{eqnarray}\label{D5}
 G^{30}_{03}\!\left( z\, \biggr|b_1, b_2,b_3\!\right)=\sum_{h=1}^3{{{\prod}'}_{j=1}^3\Gamma(b_j-b_h)} \ z^{b_h}\times {}_0F_2[1+b_h-b_1,\dots,*,\dots1+b_h-b_3;-z]
\end{eqnarray}where $ {}_0F_2$ is a generalized hypergeometric series in $z$, with \cite{Erdelyi1953},
\begin{equation}\label{D6}
{}_0F_2[1+b_h(q_i)-b_1(q_i),\dots,*, \dots,1+b_h(q_i)-b_3(q_i); -z]=1+O_{ih}\,z+{\cal{O}}(z^2)\,,
\end{equation}where in (\ref{D5}), the prime on top of the product indicates that the pole value $b_j=b_h$ is omitted, and likewise the asterisk in the ${}_0F_2$ generalized hypergeometric series (\ref{D6}), indicates that the value $b_j=b_h$ is omitted. This shows that for $|z_i|<1$, the last expression above admits the expansion,
\begin{equation*}
\pm  (-{4\pi^2 m^2\over E^2})^N\biggl\langle\prod_{i=1}^N\, [1-i(-1)^{q_i}]\,\sum_{h=1}^3A^h_i\,z_i^{b_h-\frac{1}{2}}\left(\,1+O_{ih}\,z_i+{\cal{O}}(z_i^2)\,\right)\biggr\rangle_{\!O_N(\mathbb{R})}\,,
\end{equation*}where (\ref{D5}) and (\ref{D6}) have been used, and where the variable $z_i$ is the full (matrix-valued) argument of the Meijer function $G^{30}_{03}$ defined in (\ref{z_i}). The numbers $A^h_i$ and $O_{ih}$ are those given in (\ref{Anumbers}) and (\ref{Onumbers}), respectively.  Averaging over $O_N({\mathbb{R}})$, the odd powers of $({\cal{OT}})_i$ vanish, but there remain contributions of order $z_i^0$, ${\sqrt{z_i}}$, $z_i$ and $z_i{\sqrt{z_i}}$, which are on the order of ${\mathbf{1}}_{3\times 3}$, $({\cal{OT}})_i^2$, $({\cal{OT}})_i^4$ and $({\cal{OT}})_i^6$ respectively, and are leading in view of the smallness of $\lambda$.

This illustrates that any order $n$ in an expansion of the Meijer function is well defined as an expansion in integer powers of matrices $[({\cal{OT}})_i]$, and this justifies the second and third expressions of this Appendix, or of (\ref{new1}) in the main text.

\section{\label{AppF} $`O_N({\mathbb{R}})$-averaged Baker-Campbell-Hausdorf'}

Upstream of (\ref{Eq:1}), and leading to the second line of (\ref{Eq:1}), one has an expression of \cite{QCD6},
\begin{eqnarray}\label{E1}
&& \int{\mathrm{d}^{(N_c^2-1)} \alpha_1} \int{\mathrm{d}^{(N_c^2-1)} \alpha_{2} } \int{\mathrm{d}^{(N_c^2-1)} \Omega_1} \int{\mathrm{d}^{(N_c^2-1)} \Omega_{2} \, e^{-i\alpha_1\cdot\Omega_I}\, e^{-i\alpha_{2}\cdot\Omega_{2}}\, \bigl\langle e^{i\alpha_1\cdot T}\ e^{i\alpha_{2}\cdot T} \bigr\rangle}\nonumber \\ && {\sqrt{i}}\,C_{N_1}\,\prod_{i=1}^{N}\int^{+\infty}_{-\infty} {{\rm{d}}}\xi_i\  \ e^{-\frac{i}{2}\,\xi_i^2}\prod_{i<j} (\xi_i-\xi_j)\, \ f_i(\xi_i)\,.
\end{eqnarray}where $f_i(\xi_i)={({\exp{[-b_i/\xi_i}] )/{\sqrt{\xi_i}}}}$, and $b_i$ as in (\ref{Meijer}).
Then ($N=D\times (N_c^2-1)$), some convenient change of integration variables such as defined after (\ref{Meijer}) allows one to rewrite the first line of (\ref{E1}) as \cite{QCD6},
\begin{equation*}
E_1^{-N}E_2^{-N}\int{\mathrm{d}^N {{\alpha}}_1}\, \int{\mathrm{d}^N {{\alpha}}_2}\,\bigl\langle e^{i{{\alpha}}_1\cdot\, {\cal{OT}}} \, e^{i{{\alpha}}_2\cdot\, {\cal{OT}}}\bigr\rangle\int{\mathrm{d}^N V_1}\, e^{-i{{{{\alpha}}_1\cdot\, V_I}\over E-p}} \int{\mathrm{d}^N V_{2} \, e^{-i{{{{\alpha}}_2\cdot\, V_2}\over E+p}} },
\end{equation*}where, as throughout the current paper, the brackets are here to mean an $O_N({\mathbb{R}})$-averaged quantity. The two exponentials of arguments ${{{\alpha}}_J\!\cdot\! {\cal{OT}}}\,,(J=1,2)$ are those leading to the two {\textit{sine}} functions that appear in the second line of (\ref{Eq:1}), once integrations over the $V_J^i\,, J=1,2$ are performed (level of expression (\ref{A11})).
\par\medskip
Now, the point is the following. Whereas the 2nd line of (\ref{E1}) displays a structure of a finite sum of finite products comprising $N$ terms each, the 1st line doesn't. This is of course due to the fact that both exponents are linear combinations of non-commuting $SU_c(3)$ generators (Cf. {\textit{Baker-Campbell-Hausdorf formulae}}), so that in general, one has,
\begin{equation}\label{E3}
e^{i{{\alpha}}_J\cdot\, {\cal{OT}}}\equiv\ e^{\,i\sum_{i=1}^N{{\alpha}}^i_J\, ({\cal{OT}})_i}\neq\prod_{i=1}^{N}\ e^{\,i{{\alpha}}^i_J\, ({\cal{OT}})_i}\ , \ J=1,2\,.
\end{equation}Now, it is a most important fact that under $O_N({\mathbb{R}})$- integration, equality is restored in (\ref{E3}). We will proceed  by inspection of the first few non-trivial orders and check,
\begin{equation}\label{E4}
\bigl\langle e^{i{{\alpha}}_J\cdot\, {\cal{OT}}}\bigr\rangle_{\varepsilon,\Theta}=\bigl\langle\ \prod_{i=1}^N \,e^{\,i{{\alpha}}^i_J\, ({\cal{OT}})_i}\bigr\rangle_{\varepsilon,\Theta} \ \ ,\ \ \ J=1,2\,.
\end{equation}

 At first non-trivial order (and besides the fact that odd powers of ${\cal{O}}$ trivially vanish under $O_N({\mathbb{R}})$-average) this is just,
\begin{equation}\label{E5}
\bigl\langle 1+{i\sum_{i=1}^N{{\alpha}}_J^i\, {\cal{OT}}_i}\bigr\rangle_{\varepsilon,\Theta}=\bigl\langle \prod_{i=1}^N\,(1+{i{{\alpha}}_J^i\, {\cal{OT}}_i})\bigr\rangle_{\varepsilon,\Theta}
\end{equation}
At next order, one has to compare (simplifying to the case of $N=2$, so as to avoid too lengthy expressions),
\begin{equation*}
\frac{i^2}{2}\bigl\langle (\alpha^1_J{\cal{OT}}_1)^2+\alpha^1_J\alpha^2_J\,({\cal{OT}}_1{\cal{OT}}_2+{\cal{OT}}_2{\cal{OT}}_1)+(\alpha^2_J{\cal{OT}}_2)^2\bigr\rangle_{\varepsilon,\Theta}
\end{equation*}to,
\begin{equation*}
\frac{i^2}{2}\bigl\langle (\alpha^1_J{\cal{OT}}_1)^2+2\alpha^1_J\alpha^2_J\,({\cal{OT}}_1{\cal{OT}}_2)+(\alpha^2_J{\cal{OT}}_2)^2\bigr\rangle_{\varepsilon,\Theta}\,,
\end{equation*}that is, to compare the crossed terms. Within the notations introduced for Equation (\ref{11}), one finds a result of $i^2\alpha_J^1\alpha_J^2\sum_{i=1}^N {\cal{T}}_i^2<a^{1i}a^{2i}>_\Theta$ for both crossed terms, and equality holds.
\par\medskip
At order 3, one has to compare longer expressions. These match if and only if,
\begin{equation}\label{E8}
\bigl\langle ({\cal{OT}})_2({\cal{OT}})_1({\cal{OT}})_2+({\cal{OT}}_2)^2{\cal{OT}}_1\bigr\rangle_{\varepsilon,\Theta}=\bigl\langle ({\cal{OT}})_1({\cal{OT}})_2({\cal{OT}})_1+{\cal{OT}}_2({\cal{OT}}_1)^2\bigr\rangle_{\varepsilon,\Theta}=0\,.
\end{equation}Now, besides the fact that, again, odd powers of ${\cal{O}}$ trivially vanish under $O_N({\mathbb{R}})$-average, it is easy to check that the sole $\varepsilon$-average guarantees (\ref{E8}).
\par
Fourth order is less trivial. Even at $N=2$, the corresponding expressions are quite cumbersome. At this order, the comparison of both sides of (\ref{E4}) is between the 2 members,
\begin{equation*}
\frac{(i^2\alpha^1_J\alpha^2_J)^2}{4!}\,\bigl\langle\left( {\cal{OT}}_1^2{\cal{OT}}_2^2+({\cal{OT}}_1{\cal{OT}}_2)^2+{\cal{OT}}_1{\cal{OT}}_2^2{\cal{OT}}_1+(1\leftrightarrow 2)\right)\bigr\rangle_{\varepsilon,\Theta}
\end{equation*}and, 
\begin{equation*}
(\frac{i^2\alpha^1_J\alpha^2_J}{2!})^2\,\bigl\langle( {\cal{OT}})_1^2({\cal{OT}})_2^2\bigr\rangle_{\varepsilon,\Theta}\,.
\end{equation*}All other terms, comprising odd powers of ${\cal{OT}}_1$ or ${\cal{OT}}_2$, give zero under full $O_N({\mathbb{R}})$-average.
\par\noindent
In the 1st member, the terms with subscripts $1$ and $2$ exchanged just amount to twice the contributions of the first 3 terms. Then, resorting to explicit calculations one finds,
\begin{eqnarray}\label{E11}
&&\bigl\langle {\cal{OT}}_1^2{\cal{OT}}_2^2\bigr\rangle_{\varepsilon,\Theta} =\bigl\langle {\cal{OT}}_1{\cal{OT}}_2^2{\cal{OT}}_1\bigr\rangle_{\varepsilon,\Theta} =\bigl\langle ({\cal{OT}}_1{\cal{OT}}_2)^2\bigr\rangle_{\varepsilon,\Theta} \cr &&  \qquad\qquad\qquad\ \, \,=\sum_{i,j=1}^N \bigl\langle(a^{1i})^2(a^{2j})^2\bigr\rangle_\Theta\,\left({\cal{T}}_i^2{\cal{T}}_j^2+d_{ijk}{\cal{T}}_i{\cal{T}}_j{\cal{T}}_k\right)
\end{eqnarray}where, in the second line the $SU_c(3)$ constants $d_{ijk}$ are defined with respect to the values of indices $i,j,k$ understood {\textit{modulo}} $8$. The two first equalities of (\ref{E11}) guarantee the equality of both sides of (\ref{E4}) at fourth order since one has now,
\begin{equation}\label{E12}
 2\times 3\times\,\frac{(i^2\alpha^1_J\alpha^2_J)^2}{4!}=(\frac{i^2\alpha^1_J\alpha^2_J}{2!})^2\,.\end{equation}Throughout this calculation use has been made of relations such as, 
 \begin{equation} \langle a^{ij}({\mathbf{\Theta}})a^{ik}({\mathbf{\Theta}})\rangle_\Theta=\delta^{jk}\,\langle (a^{ij})^2({\mathbf{\Theta}})\rangle_\Theta\,,\end{equation} met already at the level of (\ref{11}), and which are easily derived out of (\ref{O}) by inserting, between rotators, the relevant number of identity resolutions, ${\mathbb{I}}_N=\sum_{i=1}^N|i><i|$.  \par\medskip
In the case of two exponentials, relevant to (\ref{E1}), exactly the same proof goes through. Calculations are even easier: At $N=2$ for example, it is immediate to see that by the discrimination of $(\alpha^1_J\alpha^2_J)^2$ into $(\alpha^1_1\alpha^2_2)^2$, at any given order, the number of terms to be considered is diminished, and one gets in the same way, 
\begin{equation}\label{E13}
\bigl\langle e^{i{{\alpha}}_1\cdot\, {\cal{OT}}}\,e^{i{{\alpha}}_2\cdot\, {\cal{OT}}}\bigr\rangle_{\varepsilon,\Theta}=\bigl\langle\ \prod_{i=1}^N \,e^{\,i{{\alpha}}^i_1\, ({\cal{OT}})_i}\,e^{\,i{{\alpha}}^i_2\, ({\cal{OT}})_i}\bigr\rangle_{\varepsilon,\Theta}\,.
\end{equation}That is, over exponentials of non-commuting $SU_c(3)$-generators, $O_N({\mathbb{R}})$-integration realizes the same simplification as {\textit{Time-Ordering}} does on non-commuting field operators,
\begin{equation}
T\left(e^{\varphi(t_1)+\varphi(t_2)}\right)=T\left(e^{\varphi(t_1)}\,e^{\varphi(t_2)}\right)\,.\end{equation} Therefore expression (\ref{Eq:1}) is legitimate, with the accompanying remark that on the right hand side, the index $i$ is not to be summed upon in expressions ${{\alpha}}^i_1\, ({\cal{OT}})_i$ and ${{\alpha}}^i_2\, ({\cal{OT}})_i$. 
\par\noindent
These are the two exponentials that, one step beyond, being integrated over $\alpha^i_J$ from $-\infty$ to $+\infty$, together with another factor of $1/\alpha^i_J$ (resulting from integration on $V_J^i$ in (\ref{A11})), yield the two {\textit{sine}} functions appearing in (\ref{Eq:1}). 
\par\medskip
Relation (\ref{E13}) extends in a straight forward manner to the case of $2n$-point Green's functions and guarantees their non-trivial and interesting overall structures, as finite sums of finite products of Meijer special functions.
\par
As will be explained elsewhere, that (\ref{E13}) has to hold true is indeed related to the structure of the Wick theorem expansion for Green's functions. This structure must be the same in either perturbative and non-perturbative cases, and (\ref{E13}) guarantees that equivalence. Alternatively, this equivalence sheds some interesting light on the deep meaning of effective locality.


\begin{thebibliography}{}
%
% and use \bibitem to create references.
%
% Format for Journal Reference
\bibitem{QCD1}
 H.M. Fried, Y. Gabellini, T. Grandou, and  Y.-M. Sheu,  Eur. Phys. J. \textbf{C65} (2010), 395.


\bibitem{QCD-II}
H.M. Fried, T. Grandou, and Y.-M. Sheu.,  Ann. Phys. {\bf{327}} (2012), 2666.


\bibitem{QCD5}
H.M. Fried, Y. Gabellini, T. Grandou, and Y.M. Sheu, 
Ann. Phys.{\textbf{ 338}} ( 2013), 107. 
 

\bibitem{QCD6} H.M. Fried, T. Grandou, and Y.-M. Sheu, Ann. Phys.{\bf{344C}} (2014) 78.

\bibitem{QCD5'}  H.M. Fried, P. H. Tsang, Y. Gabellini,  T. Grandou, and Y.-M. Sheu, Ann.
Phys. 359 (2015)~1.

\bibitem{Ferrante2011}
 D.D. Ferrante, G.S. Guralnik,  Z. Guralnik  and C. Pehlevan C (2011), {BROWN-HET-1611}


\bibitem{deTeramond2012}
 G.F. de T\'eramond and S.J. Brodsky, (2012), arXiv:1203.4025 [hep-ph] ; G.F. de Teramond, S.J. Brodsky, and H.G. Dosch (2014)
  %``Hadron Spectroscopy and Dynamics from Light-Front Holography and Conformal Symmetry,''
  arXiv:1401.5531 [hep-ph].


\bibitem{tg} T. Grandou, EPL {\bf{107}} (2014),11001.


\bibitem{Nieuwenhuizen}  G.C. Nayak and P. van Nieuwenhuizen, Phys. Rev. D
{\bf{71}} (2005) 125001; G.C. Nayak, Phys. Rev. D{\bf{72}} (2005) 125010.


\bibitem{Cooper} F. Cooper, J.F. Dawson and B. Mihaila,
Phys. Rev. D {\bf{78}} (2008) 117901.


\bibitem{Dmitrasinovic}V. Dmitrasinovic Phys. Lett. B {\bf{499}} (2001), 135.


\bibitem{prep}
H.M. Fried, T. Grandou, and R. Hofman, work in progress.


\bibitem{Bali}  G.S. Bali, Phys. Rev. D {\bf{62}} (2000)114503;  Nucl.Phys.Proc.Suppl. {\bf{83}} (2000) 422; S. Delgar, Phys. Rev. D {\bf{62}} (2000) 034509; Nucl.Phys.Proc.Suppl. {\bf{73}} (1999) 587.


\bibitem{MIT} K. Johnson and C.B. Thorn, Phys. Rev. D {\bf{13}} (1976)1934; T.H. Hansson, Phys. Lett. B {\bf{166}} (1986) 343.


\bibitem{Dosch}H.G. Dosch, and Y.A. Simonov, Phys. Lett. B {\bf{205}} (1988) 339.


\bibitem{BMuller} B. M\"uller, private communication.


\bibitem{Mehta1967}
 M. L. Mehta, \textit{Random Matrices}, Academic Press, 1967.

\bibitem{Yndurain} F.J. Yndurain, {\textit{Quantum Chromodynamics}}, Texts and Monographs in Physics, Springer-Verlag, 1983.

\bibitem{Anderson et al.} T.W. Anderson, I. Olkin, and L.G. Underhill (1985), Econometric Workshop, Technical Report N06, Stanford University, California.

\bibitem{Anirban} A. Kundu, {\textit{Elementary Discussions on Group Theory}}, Calcutta University PG-I and PG-II.

\bibitem{Close} F. Close, {\textit{An Introduction to Quarks and Partons}}, Academic Press, New York, 1979.

%\bibitem{Cooper} F. Cooper, J.F. Dawson and B. Mihaila, arXiv:0811.3905v2[hep-ph].

\bibitem{AWipf} A. Wipf, private communication.


\bibitem{tg'}
T. Grandou, Work in progress.


\bibitem{Halpern1977a}
M. B. Halpern, (1977), Phys. Rev. D\textbf{16} 1798.


\bibitem{Halpern1977b}
M. B. Halpern., (1977), Phys. Rev. D\textbf{16} 3515.


\bibitem{Reinhardt} H. Reinhardt, K. Langfeld and L. v. Smekal, Phys. Lett. B{\bf{300}} (1993) 11; H. Reinhardt, "Dual description of QCD", (1996), arXiv:hep-th/9608191v1.


\bibitem{pomme}
A. Apelblat, \textit{Table of Definite and Infinite Integrals}, Elsevier Science Ltd., 1983, p.26, formula~68.

\bibitem{Erdelyi1953}
A. Erdelyi, \textit{Higher Transcendental Functions}, McGraw-Hill, 1953, Volume I, pp.209-212.


\bibitem{Erdelyi1954}
A. Erdelyi, \textit{Tables of Integral Transforms}, McGraw-Hill 1954, Volume II.







%\bibitem{PeterTsang} H. M. Fried, P.H. Tsang, Y. Gabellini, T. Grandou and Y.-M. Sheu, arXiv:1412.2072 [hep-th]








%\bibitem{Nesterenko} A.V. Nesterenko, Int. J. of Mod. Phys. A, {\b{18}} (2003) 5475.


%\bibitem{BS2009}
%S.J. Brodsky and R. Shrock, Phys. Lett. B {\bf{666}} (2010) 300.


%\bibitem{Trento}
%G. Van Baalen, D. Kreimer, D. Uminsky and K. Yeats, Ann. Phys. {\bf{325}} (2008) 95.


%\bibitem{Casher} A. Casher, Phys. Lett. B {\bf{83}} (1979) 395.


%\bibitem{Luscher} M. Luscher, K. Symanzik, P. Weisz, Nucl. Phys. B {\bf{173}} (1980) 365; M. Luscher, Nucl. Phys. B {\bf{180}} (1981) 317.


%\bibitem{Wilczek} F. Wilczek, Ann. H. Poincar\'e {\bf{4}}, Suppl. 1 (2003) S211.


\end{thebibliography}
\end{document}